\documentclass[preprint,superscriptaddress,amsmath,amssymb,floatfix, longbibliography, nofootinbib]{revtex4-1}
\usepackage{graphicx}
\usepackage{float}
\usepackage{latexsym,amsmath,amssymb,bm,euscript,multirow,makecell,slashed}
\usepackage{color}
\usepackage{wasysym}
\usepackage{epstopdf}
\usepackage[colorlinks=true,linkcolor=blue,citecolor=blue]{hyperref}
\usepackage{type1cm}
\usepackage{appendix} 

\usepackage{color}
\definecolor{darkblue}{rgb}{0.,0.,0.4}
\definecolor{darkred}{rgb}{0.5,0.,0.}
\definecolor{BlueViolet}{RGB}{138,43,226}
\definecolor{SkyBlue}{RGB}{30,144,255}
\definecolor{DarkGreen}{RGB}{0,100,0}

\begin{document}
\preprint{DESY 21-124}
\title{Conformal bootstrap bounds for the $U(1)$ Dirac spin liquid and $N=7$ Stiefel liquid
}

\author{Yin-Chen He}
\email{yinchenhe@perimeterinstitute.ca}
\affiliation{Perimeter Institute for Theoretical Physics, Waterloo, Ontario N2L 2Y5, Canada}
\author{Junchen Rong}
\email{junchen.rong@desy.de}
\affiliation{Deutsches Elektronen-Synchrotron DESY, Notkestra{\ss}e 85, 22607 Hamburg, Germany}
\author{Ning Su}
\email{suning1985@gmail.com}
\affiliation{Department of Physics, University of Pisa, I-56127 Pisa, Italy
}

\begin{abstract}
    We apply the conformal bootstrap technique to study the $U(1)$ Dirac spin liquid (i.e. $N_f=4$ QED$_3$) and the newly proposed $N=7$ Stiefel liquid (i.e. a conjectured 3d non-Lagrangian CFT without supersymmetry).
 For the $N_f=4$ QED$_3$,  we focus on  the monopole operator and ($SU(4)$ adjoint) fermion bilinear operator.
 We bootstrap their  single correlators as well as the mixed correlators between them.
We first discuss the bootstrap kinks from single correlators.
Some exponents of these bootstrap kinks are close to the expected values of QED$_3$, but we provide clear evidence that they should not be identified as the QED$_3$. 
By requiring the critical phase to be stable on the triangular and the kagome  lattice, we obtain rigorous numerical bounds for the $U(1)$ Dirac spin liquid and the Stiefel liquid. 
For the triangular and kagome Dirac spin liquid, the rigorous lower bounds of the monopole operator's scaling dimension are $1.046$ and $1.105$, respectively.
These bounds are consistent with the latest Monte Carlo results.
\end{abstract}

\maketitle
\tableofcontents
\section{Introduction}

A frontier of modern condensed matter research is to explore exotic quantum matter with long-range quantum entanglement.
Such long-range entangled phases include topological phases described by topological quantum field theories (TQFTs)~\cite{Wentopo} and critical phases described by (self-organized) conformal field theories (CFTs), i.e., CFTs without relevant singlet operators.
Compared to topological phases, critical phases are poorly understood both on the formal side of quantum field theories and on the practical side of the condensed matter realizations.
In recent years, the conformal bootstrap has become a  powerful tool to study CFTs in generic space-time dimensions~\cite{Rychkov:2009ij,ElShowk:2012ht,Kos:2013tga,Kos:2014bka,El-Showk:2014dwa,Kos:2015mba,Kos:2016ysd,Simmons-Duffin:2016wlq,Rong:2018okz,Atanasov:2018kqw,Iliesiu2016Bootstrapping,Iliesiu2018Bootstrapping,Chester:2019ifh,Chester2020O3} (see a review~\cite{Poland:2018epd}).
It produced critical exponents of $3d$ Ising~\cite{Kos:2014bka} and $O(2)$ Wilson-Fisher~\cite{Chester:2019ifh} with the world record precision, and importantly, has solved the long-standing inconsistency between Monte-Carlo simulations and experiments  of $O(2)$ Wilson-Fisher~\cite{Chester:2019ifh} as well as the cubic instability of $O(3)$ Wilson-Fisher~\cite{Chester2020O3}. 
It will be interesting to extend the success of conformal bootstrap on classical condensed matter to the frontiers of quantum matter. 

One interesting critical quantum phase is called the $U(1)$ Dirac spin liquid (DSL)~\cite{Affleck1988, Wen1995, Hastings2000, Hermele2005, Hermele2008,Song2018,Song2018a}, which is likely to be realized in several theoretical models~\cite{ Ran2007,Iqbal2013,Iqbal2016,He2017, Hu2019} (e.g. kagome and triangular spin-$1/2$ quantum magnets) as well as materials. 
Theoretically, the DSL is described by a $N_f=4$  QED$_3$ theory.
A widely believed scenario is that this QED$_3$ theory in the infared will flow into an interacting CFT with the global symmetry $\frac{SO(6)\times SO(2)}{Z_2}\times CPT$.
Here $SO(6)\sim SU(4)$ corresponds to the flavor rotation symmetry of four (2-component) Dirac fermions, while $SO(2)\cong U(1)$ is the flux conservation symmetry of the $U(1)$ gauge field. 
There is numerical evidence from Monte Carlo simulations supporting the CFT scenario~\cite{Karthik2015, Karthik2016,Karthik2019}. 
However, it is challenging for Monte Carlo to distinguish the true CFT behavior from the pseudo-critical (i.e. walking) behavior caused by the fixed points collision~\cite{wang2017deconfined,gorbenko2018walking,gorbenko2018walking2} (see \cite{Kaveh2005,Braun2014,Herbut_fixedanni,Lorenzo2016,Giombi2015} for the study of QED$_3$ in specific). 
It is important to prove or disprove whether the $N_f=4$ QED$_3$ is conformal using a more rigorous approach, such as the conformal bootstrap.

Besides showing the $N_f=4$ QED$_3$ describes a true CFT, it is also crucial to know scaling dimensions of certain operators in order to determine the fate of the DSL.  
This is because in a condensed matter realization, the system  typically has a lower UV symmetry compared to the full IR symmetry $\frac{SO(6)\times SO(2)}{Z_2}\times CPT$.
Operators that are non-trivial under the full IR symmetry could be singlet under the microscopic UV symmetry~\cite{Song2018,Song2018a}.
If such operators are relevant, they will destabilize the DSL.
In other words, the DSL will not be a stable critical phase, instead it will correspond to a critical or multi-critical point~\cite{Jian2018}.
Calculating accurate scaling dimensions of these operators is another important task to understand the DSL in the condensed matter system. 

Conformal bootstrap utilizes the intrinsic self-consistency relations (i.e. crossing symmetry) of CFT correlation functions without resorting to a specific Lagrangian~\cite{Poland:2018epd}, making it an ideal tool to study CFTs with no renormalizable Lagrangian descriptions. 
The existence of CFTs without Lagrangian descriptions as their UV completion (non-Lagrangian CFTs) is known in supersymmetric gauge theories (see, for example, Refs.~\cite{Garcia2015, Beem2016, Gukov2017}), with the most famous examples being the $\mathcal{N}=2$ supersymmetric Argyres-Douglas theories ~\cite{argyres1995new}. The Argyres-Douglas theories were long believed to have no UV Lagrangian descriptions, until a recent discovery of their $\mathcal{N}=1$ Lagrangian \cite{Maruyoshi:2016aim}.
Recently, a family of 3d non-Lagrangian CFTs without supersymmetry, dubbed Stiefel liquids, was conjectured~\cite{zou2021}. 
The Stiefel liquids can be viewed as the 3d version of the well-known 2d Wess-Zumino-Witten (WZW) CFTs. 
This is defined by a 3d non-linear sigma model on the Stiefel manifold $SO(N)/SO(4)$, supplemented with a quantized WZW term at level $k$.
The Stiefel liquids indeed naturally generalize the $SO(5)$ deconfined phase transition $(N=5,k=1)$~\cite{deccp,deccplong,Nahum_2015_Emergent,wang2017deconfined} and the aforementioned DSL $(N=6,k=1)$ to an infinite family of CFTs labeled by $(N\ge 5, k\neq 0)$. 
Unlike in 2d, there exist no direct renormalization group (RG) flow from the 3d WZW Lagrangian (which describes the symmetry breaking phase) to the Stiefel liquid fixed point (which describes the conformal phase). A phase diagram can be seen in~\cite{zou2021}. 
In this sense, the WZW descriptions are not Lagrangian UV completions of the Stiefel liquids. 
The Stiefel liquids with $N\ge 7$ are conjectured to be non-Lagrangian, due to the lack of UV completions using renormalizable Lagrangian descriptions. 
The full IR symmetry of Stiefel liquids is $SO(N)\times SO(N-4)\times CPT$~\footnote{For even $N$, the precise IR symmetry should be $\frac{SO(N)\times SO(N-4)}{Z_2}\times CPT$. This subtlety, nevertheless, may not be important for the conformal bootstrap calculation.}, and all the singlet operators under the full IR symmetry are irrelevant.
This information shall provide a good starting point to search for Stiefel liquids using conformal bootstrap.

In the condensed matter system, the $N=7$ Stiefel liquid could emerge from the intertwinement/competition between the non-coplanar magnetic order and valence bond 
solid.
This nicely generalizes the physical picture of the $SO(5)$ deconfined phase transition (i.e. intertwinement/competition between the collinear magnetic order and valence bond solid) and the DSL (i.e. intertwinement/competition between the non-collinear magnetic order and valence bond solid).
A condensed matter realization of the $N=7$ Stiefel liquid will also face the problem that the UV symmetry is much smaller than the full IR symmetry.
To determine whether the $N=7$ Stiefel liquid could be a critical phase in condensed matter system, it is important to determine if there exist relevant operators that are singlet under the UV symmetry~\footnote{For the $SO(5)$ deconfined phase transition (i.e. $(N=5,k=1)$ Stiefel liquid), the UV symmetry of its typical realization is the $SO(3)\times SO(2)$. Under this UV symmetry, there indeed exists a relevant operator, making the $(N=5,k=1)$ Stiefel liquid to be a critical point rather than a critical phase in most condensed matter systems. }.

Therefore, there are several interesting questions regarding the DSL and Stiefel liquids for the conformal bootstrap to tackle: 1) Are their effective theories true CFTs in the IR? 2) Are they quantum critical phases or quantum critical points in condensed matter systems? 3) What are the values of experimental measurable critical exponents?
 It could be a long journey to solve these challenging questions, and in this paper we will use conformal bootstrap to address a simple question: if the DSL and $N=7$ Stiefel liquid are critical phases in condensed matter systems, what are the constraints for the experimentally measurable critical exponents?
 These rigorous constraints could be used to exclude possible candidate models and materials of the DSL and Stiefel liquid in the future.
 
 The paper is organized as follows.
 We will start by studying the DSL in 
Sec.~\ref{sec:DSL}.
The DSL and $N_f=4$ QED$_3$ will be used interchangeably in this paper.
In Sec.~\ref{sec:DSL_setup} we will give a brief overview about the known results and the setup of the bootstrap calculation  of the DSL.
In Sec.~\ref{sec:kinks} we will discuss bootstrap kinks from single correlators of both the fermion bilinear operator and the monopole operator. 
Some exponents of these kinks are close to the expected values of the $N_f=4$ QED$_3$, so it is tempting to identify them as the QED$_3$.
We, however, provide clear evidence that these kinks should not be identified as the QED$_3$.
Sec.~\ref{sec:DSL_bounds} will report the numerical bounds from the mixed correlator bootstrap between the monopole and fermion bilinear operator. 
The numerical bounds obtained here are consistent with the latest Monte Carlo simulation of the $N_f=4$ QED$_3$~\cite{Karthik2015, Karthik2016,Karthik2019}.
Sec.~\ref{sec:SL} will focus on the Stiefel liquids, in specific, we will provide numerical bounds for the $N=7$ Stiefel liquid to be a stable critical phase on the triangular and kagome lattice.
We will conclude in Sec.~\ref{sec:summary}.
All the numerical results are calculated with $\Lambda=27$ (the number of derivatives included in the numerics).

\section{Dirac spin liquid}\label{sec:DSL}

\subsection{Overview} \label{sec:DSL_setup}

The DSL is described by the $N_f=4$ QED$_3$,
\begin{equation}
\mathcal{L}=\sum_{i=1}^4\bar\psi_i i\slashed D_A\psi_i+\frac{1}{4e^2}f_{\mu\nu}f^{\mu\nu}.
\end{equation}
It has a global symmetry $\frac{SO(6)\times SO(2)}{Z_2}\times CPT$, where the fermion bilinear operator (denoted by $a$) is the $SO(6)\sim SU(4)$ adjoint and $SO(2)$ singlet, while the lowest weight monopole operator (denoted by $\mathcal M_{2\pi}$) is the bi-vector of $SO(6)$ and $SO(2)$.
A natural idea to study the DSL is to bootstrap the four-point correlation functions of the fermion bilinear  and the monopole operator. 
The single correlator of either the fermion bilinear $\langle a\, a\, a\, a\rangle$ or the monopole operator $\langle \mathcal M_{2\pi} \mathcal M_{2\pi}\mathcal M_{2\pi}\mathcal M_{2\pi}\rangle $ has been explored before~\cite{Nakayama2018Bootstrap,Chester2016towards,li2018solving}, in this paper we will also study the mixed correlators of these two operators, $\langle a\, a\, \mathcal M_{2\pi}\mathcal M_{2\pi}\rangle$, $\langle a\,  \mathcal M_{2\pi}\, a\, \mathcal M_{2\pi}\rangle$.

The OPEs of the fermion bilinear ($a$) and monopole operator ($\mathcal M_{2\pi}$) are,
\begin{eqnarray}
% a \times a &= S^+ + Adj^+ +A\bar{A}^+ + S\bar{S}^+ + Adj^- + S\bar{A}^-+A\bar{S}^-, \\
a \times a &=& (S, S)^+ + (A,S)^+ +(T,S)^+ + (84,S)^+ + (A, S)^- + (45+\overline{45},S)^-, \\
\mathcal M_{2\pi} \times \mathcal M_{2\pi} &=& (S, S)^+ + (S,T)^+ + (T,S)^+ + (T,T)^+ + (A, A)^+ \nonumber \\ &&+ (A, S)^- + (S, A)^- + (T, A)^- + (A, T)^-, \\
a \times \mathcal M_{2\pi} & =& (V, V)^\pm + (64, V)^\pm + (10+\overline{10}, V)^\pm .
\end{eqnarray}
As always the superscript $+$/$-$ denotes the even/odd spin of operators appearing in the OPE.
Here we use a notation $(Rep_{SO(6)}, Rep_{SO(2)})$ to denote the representation under the global symmetry $\frac{SO(6)\times SO(2)}{Z_2}$. $S, V, T, A$ correspond to the singlet, vector, rank-2 symmetric traceless tensor, and rank-2 anti-symmetric tensor.
For other representations we use the conventional notation, namely denoting the representation by its dimension.
One shall note that even though the $SO(2)$ anti-symmetric rank-2 tensor (i.e. $A$) is the $SO(2)$ singlet, it is important to keep the distinction in order to keep track of the parity symmetry.
The parity symmetry acts trivially in the $SO(6)$ subspace, but it anti-commutes with the $SO(2)$ rotation, namely it acts as the Pauli matrix $\sigma^z$  in $SO(2)$ space~\footnote{It can be viewed as the improper $Z_2$ rotation if the $SO(2)$ is enhanced to $O(2)$.}~\cite{Chester2016towards,Song2018,zou2021}. 
Therefore, the operator in the $(Rep_{SO(6)}, A)$ representation will be parity odd, while the operator in the $(Rep_{SO(6)}, S)$ representation will be parity even.
In this notation $a$ and $\mathcal M_{2\pi}$ are in the representations $(A, A)$ and $(V, V)$, respectively.

The latest Monte-Carlo simulation on the lattice QED$_3$ gives~\cite{Karthik2015, Karthik2016,Karthik2019}
\begin{equation}
  \Delta_{\mathcal M_{2\pi}}=1.26(8), \quad \Delta_a=1.4(2).  
\end{equation}
Large-$N_f$ results are also available for several operators.
The lowest weight monopole and fermion bilinear have~\cite{Chester_2016,Dyer2013monopole},
\begin{eqnarray}
\Delta_{\mathcal M_{2\pi}} \approx  0.265 N_f-0.0383 \approx 1.02 , \quad 
\Delta_{a} = 2 - \frac{64}{3 \pi^2 N_f} \approx 1.46.
\end{eqnarray}

The lowest scalars in the channel $(S, S), (A,S), (T,S), (84,S)$ are four-fermion operators, their large-$N_f$ scaling dimensions are~\cite{Chester_2016,Jian2018,Xu2008},
\begin{eqnarray}
\Delta_{(S, S)}&=&4 + \frac{64 (2 - \sqrt{7})}{3 \pi^2 N_f} \approx 3.65, \\ 
\Delta_{(A, S)}&=&4 + \frac{4 + 8 (25 - \sqrt{2317})}{3 \pi^2 N_f} \approx 2.44, \\ 
\Delta_{(T,S)}&=& 4 - \frac{64}{\pi^2 N_f}\approx 2.38, \\ 
\Delta_{(84,S)}&=& 4 + \frac{64}{3\pi^2 N_f}\approx 4.54.
\end{eqnarray}
In the monopole sector, an important operator is the lowest scalar in the $(T, T)$ channel.
This operator is the lowest weight $4\pi$ monopole (in Ref.~\cite{Dyer2013monopole,Chester2016towards} it is called $q=1$ monopole), and its large-$N_f$ scaling dimension is,
\begin{equation}
 \Delta_{(T, T)} \approx    0.673 N_f-0.194\approx 2.5.
\end{equation}
Other scalar monopoles appearing in the OPE are in the representations $(S, T)$, $(64, V)$, $(10+\overline{10}, V)$. 
The first one corresponds to $4\pi$ monopoles, while the last three are $2\pi$ monopoles.
For completeness, the large-$N_f$ scaling dimensions (up to the order of $N_f$) of the lowest scalars in these channels are $\Delta_{(S,T)}\approx 4.23$, $\Delta_{(64,V)}\approx \Delta_{(10,V)}\approx\Delta_{(\overline{10},V)}\approx 3.85$.

It is important to know accurate scaling dimensions of various operators listed above.
Firstly, in order to have a true CFT in the IR, it is necessary that $\Delta_{(S,S)}>3$, otherwise the conformal QED$_3$ fixed point will disappear through the fixed point collision mechanism~\cite{gorbenko2018walking,gorbenko2018walking2}. 
Secondly, materials and theoretical spin models that realize DSL have a UV symmetry which is much smaller than the IR global symmetry (i.e. $\frac{SO(6)\times SO(2)}{Z_2}\times CPT$) of the QED$_3$. 
So if the DSL is a stable quantum critical phase of matter, as opposed to a quantum critical or multi-critical point, all operators that are singlet under the UV symmetry have to be irrelevant. 
The lattice quantum numbers of these operators were thoroughly analyzed in Ref.~\cite{Song2018,Song2018a}, and it was found that: 1) For the triangular lattice spin-$1/2$ magnets, $(T, S)$, $(84, S)$ are singlets under the UV symmetry~\footnote{Indeed these two operators are UV singlets in any lattice QED$_3$ gauge model. 
So if the $N_f=4$ QED$_3$ is found to be stable in a lattice gauge model without fine-tuning, these two operators shall be irrelevant.}; 
2) For the kagome lattice spin-$1/2$ magnets, $(T, S)$, $(84, S)$, $(T, T)$, $(64, V)$ are singlets under the UV symmetry.
So the relevance or irrelevance of these operators crucially determine the fate of the DSL even if the $N_f=4$ QED$_3$ itself flows to a CFT.
It is worth noting that the large-$N_f$ results have $\Delta_{(T,S)}, \Delta_{(T,T)}<3$, so it is of priority to determine their accurate scaling dimensions.
At last, we remark that the scaling dimensions of the lowest weight monopole  $\Delta_{\mathcal M_{2\pi}}$ and fermion bilinear $\Delta_a$ can in principle be measured experimentally in DSL materials.

\subsection{Bootstrap kinks from the single correlators} \label{sec:kinks}

It is well known that the $O(N)$ Wilson-Fisher CFTs are located at the kinks of numerical bootstrap bounds of $O(N)$ symmetric CFTs~\cite{ElShowk:2012ht,Kos:2013tga}.
However, there is no a priori reason that bootstrap kinks should correspond to CFTs, not to mention known CFTs. 
In the past few years quite a few bootstrap kinks were found in numerical calculations for various global symmetries~\cite{Ohtsuki:thesis,Rong:2017cow,li2018solving,Stergiou:2019dcv,Paulos:2019fkw,he2020nonwilsonfisher,li2020searching,Reehorst2020,he2021scalarQED,vichiscalar}.
Many of these bootstrap kinks are not yet identified as any known CFTs.
The bootstrap bounds from the single correlator $\langle a\, a\, a\, a\rangle$ and $\langle \mathcal M_{2\pi} \mathcal M_{2\pi}\mathcal M_{2\pi}\mathcal M_{2\pi}\rangle $ also have kinks, and we will show that these kinks should not be identified as the DSL.

\begin{figure}
    \centering
    \includegraphics[width=0.9\textwidth]{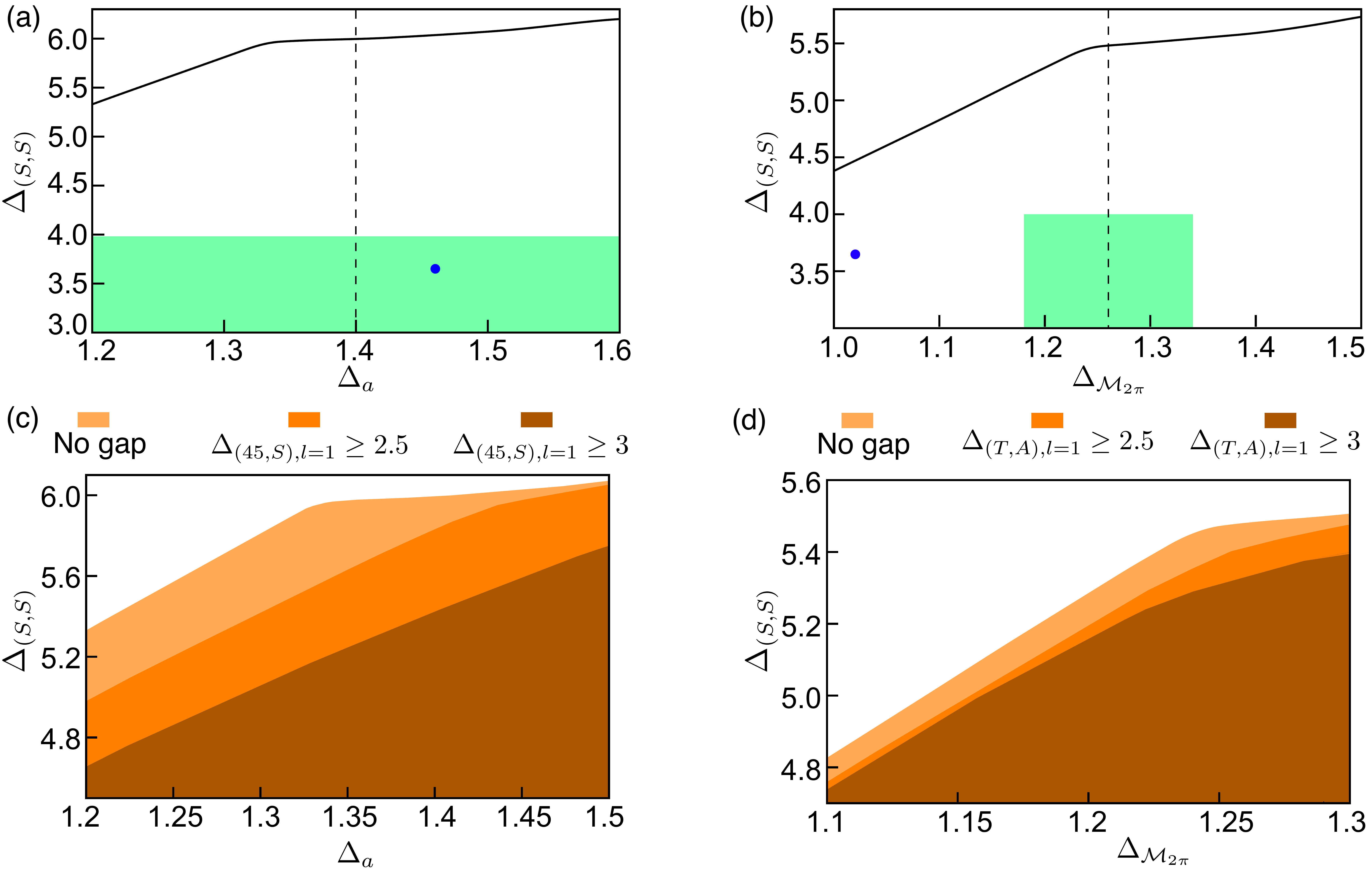}
    \caption{(a-b) Numerical bounds (black curve)  of the lowest scalar in the singlet  channel $\Delta_{(S,S)}$ (a) from the single correlator of the $SO(6)$ adjoint (i.e. fermion bilinear operator denoted by $a$) and (b) from  the single correlator of the $SO(6)\times SO(2)$ bi-vector (i.e. monopole operator denoted by $\mathcal M_{2\pi}$). The blue circles correspond to the large-$N_f$ results of the $N_f=4$ QED$_3$. The dashed lines are the Monte Carlo estimates for the $N_f=4$ QED$_3$, namely $\Delta_a=1.4(2)$ and $\Delta_{\mathcal M_{2\pi}}=1.26(8)$~\cite{Karthik2015, Karthik2016,Karthik2019}. 
    The region in green is where the $N_f=4$ QED$_3$ likely lives in.  (c-d): The zoomed in numerical bounds $\Delta_{(S,S)}$ with various gap assumptions imposed. The kinks disappearing under a mild gap shows that they should not be identified as the $N_f=4$ QED$_3$. Here we take $\Lambda=27$.
    }
    \label{fig:singlet}
\end{figure}

As shown in Fig.~\ref{fig:singlet}(a)-(b), there are kinks on the bootstrap bounds of $\Delta_{(S,S)}$ (i.e. lowest lying singlet) from single correlators. 
Somewhat curiously, the $x$-coordinates ($\Delta_a$ or $\Delta_{\mathcal M_{2\pi}}$) of the kinks are close to the best estimates (i.e. dashed line) of Monte Carlo simulations of the $N_f=4$ QED$_3$.
Based on this observation, Ref.~\cite{li2018solving} conjectured that the kink in Fig.~\ref{fig:singlet}(a) is the $N_f=4$ QED$_3$ (the kink in Fig.~\ref{fig:singlet}(b) is new here).
However, one should not overlook the fact that the $y$-coordinates of these kinks are much larger than what are expected for the QED$_3$, as $\Delta_{S,S}$ in any QED$_3$ theory will be smaller than $4$.
This large discrepancy should not be ascribed to the numerical convergence, as there is no indication that an infinite $\Lambda$ will bring the bounds of $\Delta_{(S,S)}$ down to $4$. 

Moreover, we find that even though we are bootstrapping the single correlators of the $SO(6)$ adjoint and $SO(6)\times SO(2)$ bi-vector, the numerical bounds of $\Delta_{(S,S)}$ are identical to the numerical bounds (of singlet scalars) from the single correlators of $SO(15)$ and $SO(12)$ vector, respectively~\footnote{ Similar phenomenon has been observed  before~\cite{Poland:2011ey,li2020searching}, and an explanation was provided in~\cite{Poland:2011ey}.}. 
This brings the possibility that the theories sitting at the numerical bounds (including the kinks) shown in Fig.~\ref{fig:singlet}(a)-(b) have enhanced symmetries. 
A way to see whether the symmetry enhancement happens is to investigate the symmetry current. 
For example, if a $SO(6)$ theory is enhanced to a $SO(15)$ theory, the lowest spin-$1$ operator in the $(45+ \overline{45})$ representation of the $SO(6)$ theory should be conserved.
Similarly, if a $SO(6)\times SO(2)$ theory is enhanced to a $SO(12)$ theory, the lowest spin-$1$ operators in the $(T, A)$ and $(A, T)$ representations of the $SO(6)\times SO(2)$ theory should be conserved\footnote{This comes from the branching of $SO(15)\rightarrow SO(6)$ for the $SO(15)$ current, namely
$SO(15) \textrm{ current} \rightarrow SO(6)  \textrm{ current} + 45 + \overline{45}$, as well as the branching of $SO(12)\rightarrow SO(6)\times SO(2)$ for the $SO(12)$ current, namely  $SO(12) \textrm{ current} \rightarrow (A, S) + (S, A) + (T, A) + (A, T)$. (Here $(A, S)$ and $(S, A)$ are the $SO(6)$ and $SO(2)$ current, respectively).}.
In contrast, for the QED$_3$ theory we have $\Delta_{(45+\overline{45},S),l=1}\approx 5+O(1/N_f)$, and $\Delta_{(T,A),l=1}\approx 4+O(1/N_f)$.
Therefore, we can add gaps in these channels to see how the numerical bounds and kinks are moving. 
As shown in Fig.~\ref{fig:singlet}(c)-(d), once a mild gap assumption is imposed in the spectrum, the numerical bounds are improved significantly, and the kinks disappear. 
This behavior is an indication of the symmetry enhancement, although a thorough study is necessary to make a firm conclusion.
Nevertheless, it is already enough to confirm that these two kinks are not the QED$_3$, because the scaling dimensions of $\Delta_{(45+\overline{45},S),l=1}$ and $\Delta_{(T,A),l=1}$ at the kinks are not consistent with the QED$_3$.
We also remark that a similar conclusion applies to other kinks from the $SU(N_f)$ adjoint single correlator discussed in Ref.~\cite{li2018solving}.

\subsection{Numerical bounds for the Dirac spin liquid} \label{sec:DSL_bounds}
\begin{figure}
    \centering
    \includegraphics[width=\textwidth]{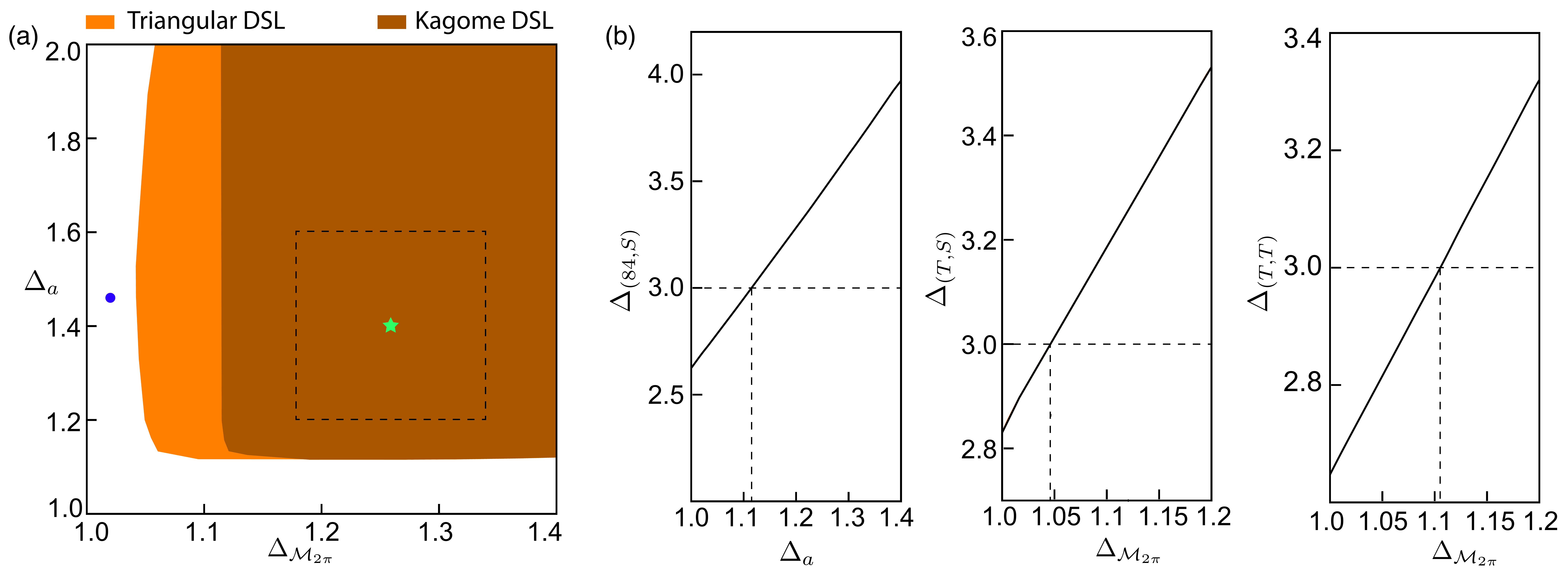}
    \caption{(a) The allowed region (shaded)  from the mixed correlator of the fermion bilinear ($a$) and monopole ($\mathcal M_{2\pi}$). For the triangular DSL, we impose the gap conditions $\Delta_{(S, S)}, \Delta_{(84,S)}, \Delta_{(T,S)},\Delta_{(V,V)'}, \Delta_{(A,A)'}>3$. 
    For the kagome DSL, we impose two more conditions $\Delta_{(T,T)},\Delta_{(64,V)}>3$. The blue circle is the large-$N_f$ result of the $N_f=4$ QED$_3$, while the green star is the estimate from the Monte Carlo simulation, with the error bars represented by the dashed line box. (b) The numerical bounds of $\Delta_{(84,S)}$, $\Delta_{(T,S)}$ and $\Delta_{(T,T)}$ from the single correlators. The last two numerical bounds were also reported in Ref.~\cite{Chester2016towards}. The allowed region is below the black curve. Here we take $\Lambda=27$.}
    \label{fig:DSL_bounds}
\end{figure}

We now turn to the numerical bounds for the DSL.
Requiring in bootstrap that operators which are singlets of the UV symmetry be irrelevant can provide lower bounds for the scaling dimension of several operators. This type of study was initiated in \cite{Nakayama:2016jhq}.
In the case of DSL, as summarized in Sec.~\ref{sec:DSL_setup}, in order to have the triangular DSL be a stable critical phase, we shall have $\Delta_{(T,S)}>3$ and $\Delta_{(84,S)}>3$.
For the kagome DSL to be stable, we have two more requirements, $\Delta_{(T,T)}>3$ and $\Delta_{(64,V)}>3$.
We thus impose these conditions in the mixed correlator (between the fermion bilinear $a$ and the monopole $\mathcal M_{2\pi}$) bootstrap calculations.
In this setup we identify OPE coefficients $\lambda_{\mathcal M_{2\pi} \mathcal M_{2\pi} a} = \lambda_{a \mathcal M_{2\pi}  \mathcal M_{2\pi} }$, and there are no ratios of OPE coefficients  to scan.
Somewhat disappointingly, unlike the mixed correlator bootstrap for the $O(N)$ Wilson-Fisher CFT~\cite{Kos:2016ysd,Kos:2015mba}, here the mixed correlator bootstrap does not produce any sharp signature for the $N_f=4$ QED$_3$.
Instead, we just get numerical bounds for $(\Delta_a,\Delta_{\mathcal M_{2\pi}})$ as shown in Fig.~\ref{fig:DSL_bounds}(a).
For the triangular DSL and kagome DSL, the allowed regions are slightly different. The large-$N_f$ result (blue circle) lies outside these regions, meaning that the large-$N_f$ monopole scaling dimension is inconsistent with DSL being a stable phase on any lattice.
While the result from the lattice Monte Carlo simulation (green star) is consistent with the DSL being stable on both lattices.

We find that numerical bounds from the mixed correlator bootstrap are almost identical to the bounds from the single correlators (Fig.~\ref{fig:DSL_bounds}(b)).
For example, from the single correlator of the fermion bilinear ($a$) we have  $\Delta_a>1.12$ if we require $\Delta_{(84,S)}>3$.
Similarly, from the single correlator of the monopole ($\mathcal M_{2\pi}$) we have $\Delta_{\mathcal M_{2\pi}}>1.046$ and  $\Delta_{\mathcal M_{2\pi}}>1.105$ if we require $\Delta_{(T,S)}>3$ and $\Delta_{(T,T)}>3$, respectively.
Other requirements such as $\Delta_{(S,S)}>3$ and $\Delta_{(64,V)}>3$ do not produce any tighter bounds. 
We have also explored other mixed correlators, including two-operator mix---$\mathcal M_{2\pi}$ with $(T, T)$ and $\mathcal M_{2\pi}$ with $(T, S)$, as well as three-operator mix between any three of $a$,  $\mathcal M_{2\pi}$, $(T, T)$, $(T, S)$.
Some of these mixed correlators could produce  bounds tighter than the single correlators (under aggressive gap assumptions), but no sharp signature of the $N_f=4$ QED$_3$ is found.

\section{$N=7$ Stiefel liquid} \label{sec:SL}

\subsection{Overview} \label{sec:SL_setup}
Stiefel liquids~\cite{zou2021} are recently proposed 3d CFTs described by 3d non-linear sigma models on the Stiefel manifold $SO(N)/SO(4)$, supplemented with a level-$k$ Wess-Zumino-Witten (WZW) term,
\begin{equation}
S[n]=\frac{1}{2g}\int d^{d+1}x\textrm{Tr}(\partial_\mu n^T\partial^\mu n) + k\cdot \textrm{WZW}.
\end{equation}
Here $n$ is a $N\times (N-4)$ matrix field, which lives on the Stiefel manifold $SO(N)/SO(4)$.
The proposal is that, there are three fixed points as one tunes the coupling constant $g$: 1) an ordered (attractive) fixed point at $g=0$, which corresponds to a spontaneous symmetry breaking fixed point with the groundstate manifold $SO(N)/SO(4)$; 2) a repulsive fixed point at $g_c$, which corresponds to the order-disorder transition; 3) a disordered (attractive) fixed point at $g\sim 1$, which is conjectured to be conformal. 
This conjectured disordered conformal fixed point is the Stiefel liquid, and we label it as SL$^{N, k}$.
The global symmetry of SL$^{N, k}$ is $SO(N)\times SO(N-4)\times CPT$ for odd $N$, and is $\frac{SO(N)\times SO(N-4)}{Z_2}\times CPT$ for even $N$.
The most important operator of SL$^{(N, k)}$ is the order parameter field $n$, which is the bi-vector of $SO(N)$ and $SO(N-4)$.

The non-linear sigma model is non-renormalizable in 3d, making it hard to study analytically.
Interestingly, SL$^{N,k}$ has a simple UV completion when $N=5,6$.
SL$^{5,k}$ is dual to a gauge theory with $N_f=2$ Dirac fermions coupled to a $USp(2k)$ gauge field, and the order parameter field can be identified as the fermion bilinear operator that is a $SO(5)$ vector.
SL$^{6,k}$ is dual to a gauge theory with $N_f=4$ Dirac fermions coupled to a $U(k)$ gauge field, and the order parameter field can be identified as the lowest weight monopole operator that is a bi-vector of $SO(6)$ and $SO(2)$.
So SL$^{6,1}$ is just the DSL we discussed in the previous section, and SL$^{5,1}$ is the $SO(5)$ deconfined phase transition~\cite{deccp,deccplong,Nahum_2015_Emergent,wang2017deconfined}.
There is no obvious gauge theory description for the SL$^{N,k}$ with $N\ge 7$, and SL$^{N\ge 7,k}$ is conjectured to be non-Lagrangian.

Based on the gauge theory description of SL$^{5,k}$ and SL$^{6,k}$, we expect that for a given $N$, the larger $k$ is, the more likely SL$^{N,k}$ will not be conformal.
It will be interesting to determine the conformal window of SL$^{N,k}$.
A natural idea to bootstrap SL$^{N,k}$ is to bootstrap the order parameter field $n$, which is the bi-vector of $SO(N)$ and $SO(N-4)$.
Below we will investigate the single correlator bootstrap of $n$, and we will focus on $N=7$.
The OPE of $n$ is the same as that of the monopole in DSL, 
\begin{equation}
\begin{aligned}
    n \times n = \quad & (S, S)^+ + (S,T)^+ + (T,S)^+ + (T,T)^+ + (A, A)^+  \\
    &+(A, S)^- + (S, A)^- + (T, A)^- + (A, T)^-.
\end{aligned}
\end{equation}
Again we are using the notation $(Rep_{SO(7)},Rep_{SO(3)})$ to denote the representation of the global symmetry of $SO(7)\times SO(3)$, with $S, A, T$ referring to the $SO(N)$ singlet, rank-2 anti-symmetric tensor, and rank-2 symmetric traceless tensor.

\subsection{Numerical results} \label{sec:SL_numerics}

\begin{figure}
    \centering
    \includegraphics[width=0.7\textwidth]{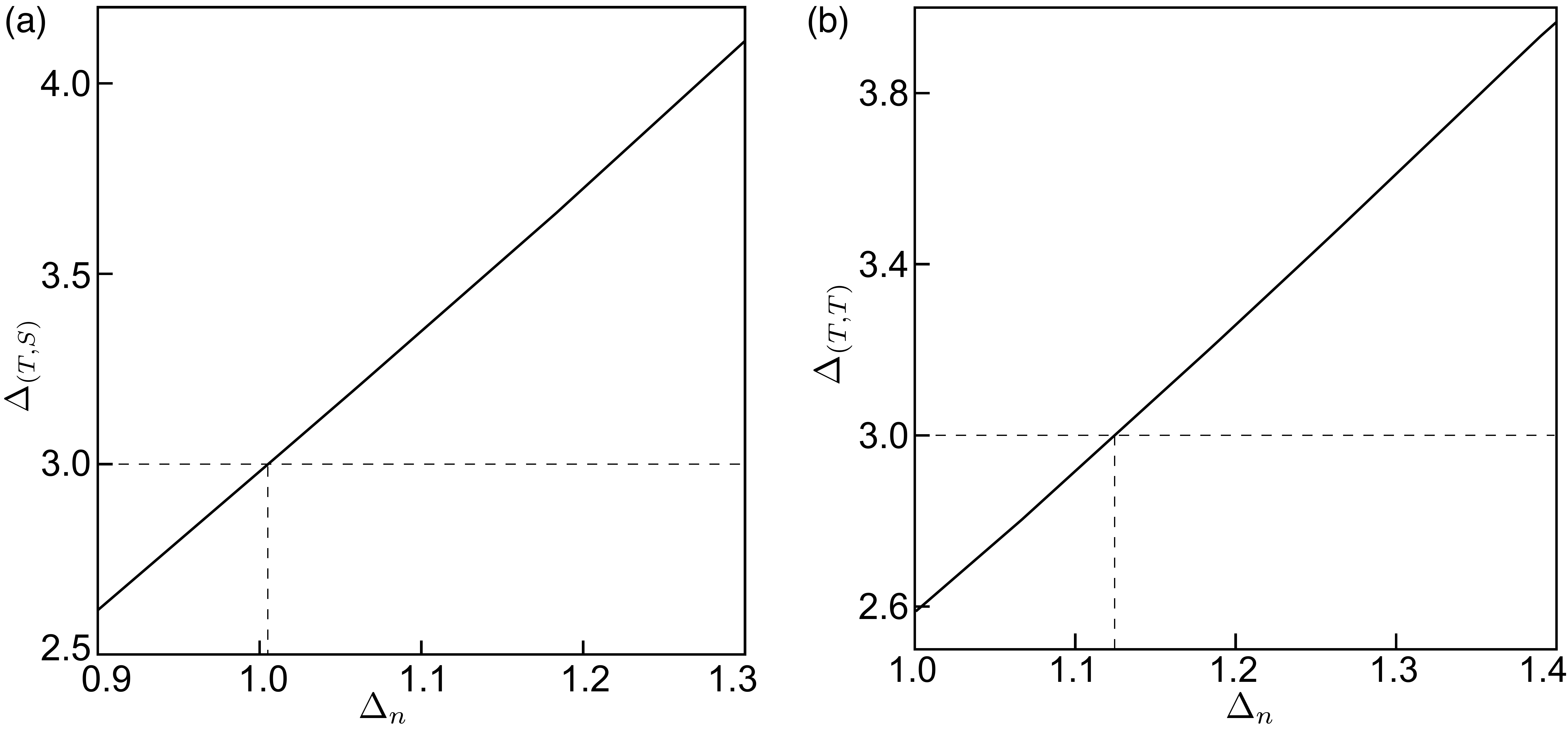}
    \caption{(a-b) Numerical bounds of $\Delta_{(T,S)}$ and $\Delta_{(T,T)}$ for the $N=7$ Stiefel liquid. The allowed region is below the black curve. Here we take $\Lambda=27$.}
    \label{fig:SL}
\end{figure}

Similar to the monopole bootstrap discussed in the previous section, the numerical bound of $\Delta_{(S,S)}$ has a kink.
Moreover, this kink also seems to have an enhanced symmetry, namely $SO(7)\times SO(3)\rightarrow SO(21)$.
So this kink is the $O(N)$ non-Wilson-Fisher kink~\cite{Ohtsuki:thesis,li2018solving,li2020searching,he2020nonwilsonfisher}, and it should not be identified as the Stiefel liquid.
We also bound other operators, but we do not find any interesting feature that can be related to the Stiefel liquid.
Nevertheless, one piece of useful information that can be extracted is the lower bound of the scaling dimension of the order parameter field. 

Possible realizations of SL$^{7,1}$ have  been proposed for the triangular and kagome quantum magnets~\cite{zou2021}.
For the proposed SL$^{7,1}$ on the triangular lattice, we shall have $\Delta_{(T,S)}>3$.
As shown in Fig.~\ref{fig:SL}(a), this requires that $\Delta_n> 1.003$.
For the proposed realization on the kagome lattice, we shall have both $\Delta_{(T,S)}>3$ and $\Delta_{(T,T)}>3$, requiring that $\Delta_n> 1.123$ as shown in Fig.~\ref{fig:SL}(b).
These bounds can also apply to SL$^{7,k}$ with $k>1$.

\section{Summary and Discussion} \label{sec:summary}

We used the conformal bootstrap to study the DSL and $N=7$ Stiefel liquid.
For the DSL, i.e. $N_f=4$ QED$_3$, we studied the single correlators of monopole ($SO(6)$, $SO(2)$ bi-vector) and fermion bilinear ($SO(6)$ adjoint), as well as their mixed correlators.
We first discuss the kinks of numerical bounds of the lowest lying singlet.
We provide clear evidence that these kinks should not be identified as the QED$_3$ theory.
We show that these kinks are likely to have enhanced symmetries, hence are the previously studied $O(15)$ and $O(12)$ non-Wilson-Fisher kinks.
By requiring the critical phase to be stable on the triangular and the kagome  lattice, we obtain rigorous numerical bounds for the scaling dimensions of certain operators of the $U(1)$ Dirac spin liquid and the Stiefel liquid. 
 
The scaling dimension of the fermion bilinear operator should be larger than $1.12$.
Furthermore, for the triangular magnets, the scaling dimension of the monopole operator should be larger $1.046$; while for the kagome magnets, the scaling dimension of the monopole operator should be larger $1.105$. 
These bounds are consistent with the latest Monte Carlo simulation of the $N_f=4$ QED$_3$.
These rigorous bounds also apply to the $U(k\ge 2)$ DSL discussed in Ref.~\cite{Vladimir2020}.

We also bootstrapped the single correlator of the $SO(7)\times SO(3)$ bi-vector, which applies to the $N=7$ Stiefel liquid with arbitrary $k$.
Similarly, we have obtained rigorous bounds assuming that the $N=7$ Stiefel liquid is a stable phase for several concrete proposed realizations.

The rigorous bounds obtained here might be useful to exclude future candidate theoretical models and materials for the DSL and $N=7$ Stiefel liquid.
It will be exciting if the conformal bootstrap can provide more accurate information of these two critical theories.
For the DSL, we have explored extensively the mixed correlators of various operators, including the fermion bilinear, $2\pi$ monopole, $4\pi$ monopole and a four-fermion operator.
We did not find any sharp signature of the DSL in these mixed correlator studies. 
For example, as shown in the paper the mixed correlator of the fermion bilinear and $2\pi$ monopole operator does not yield a tighter bound compared to the single correlator bound. 
Results of other operators mix are similar and are not illuminating to discuss in detail.
It is worth remarking that there was an expectation that the monopole  could be used to distinguish QED$_3$ from its cousin QCD$_3$ theories. 
This expectation, however, is incorrect, because the QCD$_3$ theories with $U(k)$ gauge fields also have monopole operators that share qualitatively similar properties (i.e. $SU(N_f)$ quantum numbers and scaling dimensions)~\cite{Dyer2013monopole} with the monopoles of QED$_3$.
Possible progress of bootstrapping QED$_3$ might be made by using the idea of decoupling operators proposed by us in Ref.~\cite{he2021scalarQED}, which will be pursued in future.

\begin{acknowledgments}
We would like to thank Chong Wang for discussions and Nikhil  Karthik for communicating Monte Carlo results of the $N_f=4$ QED$_3$. 
Research at Perimeter Institute is supported in part by the Government of Canada through the Department of Innovation, Science and Industry Canada and by the Province of Ontario through the Ministry of Colleges and Universities. 
This project has received funding from the European Research Council (ERC) under the European Union’s Horizon 2020 research and innovation programme (grant agreement no. 758903). 
The work of J.R. is supported by the DFG through the Emmy Noether research group ``The Conformal
Bootstrap Program'' project number 400570283. 
The numerics is solved using SDPB program \cite{Simmons-Duffin:2015qma} and
simpleboot (https://gitlab.com/bootstrapcollaboration/simpleboot). 
The computations in this paper were run on the Symmetry cluster of Perimeter institute.
\end{acknowledgments}

\bibliography{ref.bib}

\end{document}